\documentclass[reprint,superscriptaddress,amsmath,amssymb,aps,prl,showkeys,nofootinbib
]{revtex4-2}

\usepackage{graphicx,color,appendix}% Include figure files
\usepackage{dcolumn}% Align table columns on decimal point
\usepackage{bm}% bold math
\DeclareMathOperator\e{e}

\begin{document}

\preprint{APS/123-QED}

\title{Effects of stickiness in quantum chaotic billiards with $n$-fold symmetry}

\normalsize

\author{R. B. do Carmo}
\thanks{Corresponding author: ricardo.carmo@ifal.edu.br}
\affiliation{Laborat\'orio de F\'isica, Instituto Federal de Alagoas, Piranhas, AL 57460-000, Brazil}

\author{T. Ara\'ujo Lima}
\thanks{tiago.araujol@ufrpe.br}
\affiliation{Departamento de F\'isica, Universidade Federal Rural de Pernambuco, Recife, PE 52171-900, Brazil}

\date{\today}

\begin{abstract}

In this work, we study a family of fully chaotic billiards that exhibits only rotational symmetries, whose geometry is based on the $C_3$ symmetry system proposed by Leyvraz, Schmit, and Seligman~(LSS) in 1996. Quantum spectral analyses are performed on billiards with symmetry $C_n$~(the billiard repeats itself under rotations of $2\pi/n$), where $n$ is the symmetry parameter. In these systems, there are subspaces of singlets~(invariant under time reversal) and doublets~(not invariant under time reversal). For the LSS billiard, it has been established both numerically and experimentally that the corresponding subspectra follow the Gaussian Orthogonal Ensemble~(GOE) statistics for singlets and the Gaussian Unitary Ensemble~(GUE) statistics for doublets. From a classical perspective, the shapes of these billiards allow certain subregions of phase space to be visited more frequently by chaotic trajectories, a phenomenon known as stickiness. We investigate the relationship between the fraction of sticky regions in classical phase space and the deviations of the energy subspectra from GOE and GUE statistics. Our results suggest the existence of correlations between the energy distributions associated with different symmetry subspaces. In addition, we discuss aspects related to the superposition of the different energy subspectra.

%\begin{description}
%use the optional argument of the \verb+\item+ command to give the category of each %item. 
%\end{description}

\end{abstract}

%\keywords{Billiards. Quantum chaos. GOE. GUE. Symmetry.}%Use showkeys class option if keyword display desired

\maketitle

%\tableofcontents

%\section{Introduction}
%\label{sec:intro}

{\it Introduction} -- Symmetries play a fundamental role in classical and quantum physics, constraining dynamics, conservation laws, and spectral properties~\cite{gre:1994,rob:1989,lau:1991,kea:1997,joy:2012,bla:2025}. In quantum systems, discrete symmetries and time-reversal invariance are particularly important, as they determine symmetry-resolved spectral statistics. Within the framework of quantum chaos, universal features of quantum spectra are understood through Random Matrix Theory~(RMT)~\cite{meh:2004,rob:2012}. According to the Berry–Tabor conjecture, typical classically integrable systems exhibit Poissonian level statistics~\cite{ber:1977}, while the Bohigas–Giannoni–Schmit~(BGS) conjecture predicts that classically chaotic systems display universal correlations described by Gaussian random-matrix ensembles~\cite{boh:1984,mul:2004,mul:2005,heu:2007,mul:2009,ull:2016,shn:2020,cad:2024}. For spinless systems, time-reversal symmetry implies Gaussian Orthogonal Ensemble~(GOE) statistics, whereas its absence leads to Gaussian Unitary Ensemble~(GUE) behavior. Billiards provide a paradigmatic setting to test these predictions, as their classical dynamics ranges from integrable to fully chaotic depending on boundary geometry~\cite{hel:1984,che:2006,sin:1970,bun:1974,men:2007,vas:2022,vas:2026,hel:1993,qui:2025,can:1998,ara:2015,bun:2001,bar:2007,die:2007,gom:2005,gon:2025}. Quantum billiards are obtained by solving the Helmholtz equation with Dirichlet boundary conditions, enabling high-precision numerical access to long spectral sequences. In a seminal work, Leyvraz, Schmit, and Seligman~(LSS)~\cite{ley:1996} showed that chaotic billiards with only threefold rotational symmetry~($C_3$) exhibit GUE statistics in symmetry-resolved doublets~(Kramers-type degeneracy), while singlets follow GOE statistics. This property arises from the fact that certain symmetry subspaces are not invariant under time reversal, although the system is invariant~\cite{li:2020,zha:2023,ara:2024}.

In this letter, we generalize this scenario by quantizing a family of billiards that are classically fully chaotic~(ergodic) and possess discrete rotational symmetry $C_n$. This system was recently introduced in~\cite{car:2026} and preserves the geometric construction proposed by LSS, in which straight boundary segments are connected by circular arcs. As a consequence, the classical dynamics exhibits regions of stickiness in phase space, i.e., regions that are more frequently visited by chaotic trajectories~\cite{zas:2007,mak:2001,arm:2004,alt:2005,sze:2005,zou:2007,liv:2012,mak:2018,san:2019,loz:2020,mak:2022,pal:2022,sal:2023,bor:2024}. The presence of stickiness and its effects on quantization were recently investigated in~\cite{loz:2021} for the ergodic lemon billiard, showing that stickiness in the classical phase space leads to nonuniversal behavior in the energy-level statistics and in measures of localization. The main goal of the current work is to numerically investigate the nonuniversal behavior of the energy-level statistics arising from the presence of stickiness combined with the rotational symmetry of the billiards. Although the billiard is fully chaotic, we observe that the singlet states deviate from GOE statistics and the doublet states deviate from GUE statistics, as a consequence of the strong presence of stickiness in the corresponding classical dynamics.

%\section{Classical $C_n$-Symmetrical Chaotic Billiards}
%\label{sec:classical}

\begin{figure}[!htpb]
 \centering
 \includegraphics[width=1.0\columnwidth]{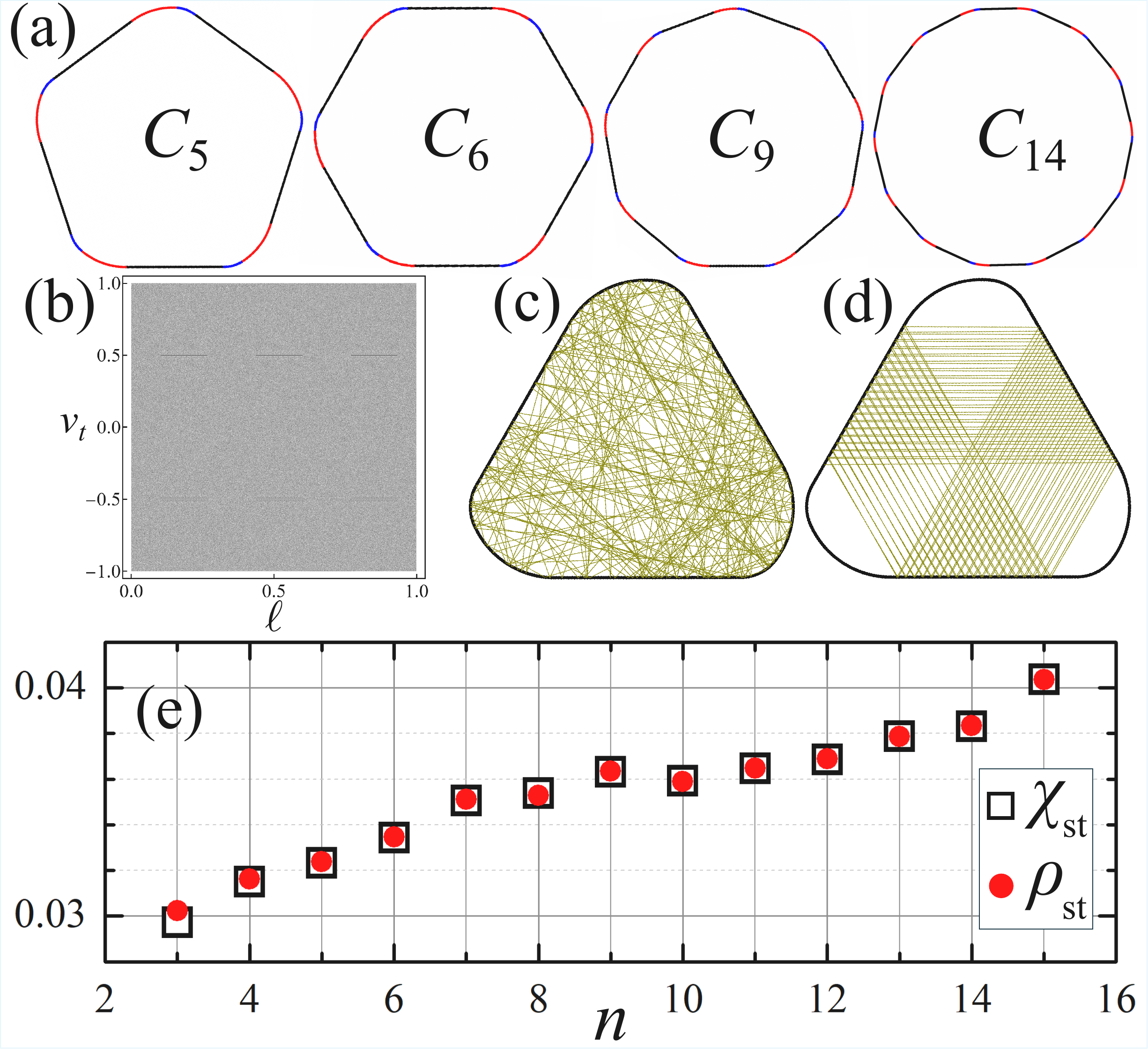}
 \caption{(a) Examples of $C_n$-SCB boundaries for several values of $n$. (b) Typical phase space for $n=3$. There are six sticky regions around $v_t \pm 0.5$. (c) Typical chaotic trajectory in real space. (d) Temporary trapping episode of stickiness. (e) The area $\chi_\text{st}$ and the hypervolume $\rho_\text{st}$ for several values of $n$.}
 \label{fig:billiards}
\end{figure}

{\it Classical $C_n$-Symmetrical Chaotic Billiards} -- In their original model proposed by LSS, an equilateral triangle is modified by replacing its vertices with circular arcs of radii $R$ and $r$ satisfying $R=2r$. This idea was generalized by considering regular polygons with $n$ sides whose vertices are replaced by two circular arcs of radii $R$ and $r$~\cite{car:2026}, red and blue arcs, respectively, in Fig.~\ref{fig:billiards}(a). Where representative billiard shapes for several values of $n$ are shown. We refer to this class as \textit{$C_n$ symmetric chaotic billiards} ($C_n$-SCB). The systems considered belong to the ergodic class and display no stability islands~\cite{alt:2008,bun:2012}. Fig.~\ref{fig:billiards}(b) shows the phase space for the $C_3$ billiard with $r=0.25$, where $\ell$ denotes the normalized arc length along the boundary and $v_t$ the tangential velocity at collision. Typical trajectories exhibit alternating dynamical regimes, illustrated in Figs.~\ref{fig:billiards}(c) and~\ref{fig:billiards}(d), where $200$ successive collisions are displayed. Periods of strongly chaotic dispersion coexist with temporary trapping episodes associated with the phenomenon of \textit{stickiness}. These trapping intervals occur near specific values of $v_t$ and are related to marginally unstable periodic orbits (MUPOs)~\cite{alt:2008,pil:2009}. In order to detect stickiness, we adopt a box-counting method recently proposed in~\cite{car:2026}. The fraction of the surface of section associated with stickiness defines $\chi_\text{st}$.

While $\chi_\text{st}$ measures an area in the two-dimensional phase plane $(\ell,v_t)$, the physically relevant quantity is the volume fraction $\rho_\text{st}$ in the four-dimensional phase space $(x,y,v_x,v_y)$ (Liouville measure). Following Meyer~\cite{mey:1986,loz:2022,ara:2024}, the two quantities are related by
\begin{equation}
\rho_\text{st} =
\frac{\chi_\text{st}}
{\chi_\text{st} + (1-\chi_\text{st})\,\langle d_\text{c} \rangle / \langle d_\text{st} \rangle } ,
\label{eq:meyer}
\end{equation}
where $\langle d_\text{st}\rangle$ and $\langle d_\text{c}\rangle$ denote the average distance between successive collisions for sticky and purely chaotic trajectories, respectively. The averages are computed from $20$ random initial conditions over trajectories of up to $2\times10^{7}$ collisions. The resulting values of $\chi_\text{st}$ and $\rho_\text{st}$ are shown in Fig.~\ref{fig:billiards}(e) for several values of $n$ and $r=0.25$, revealing a clear increase of the sticky phase-space fraction as the symmetry order grows. More details about the geometric construction of the billiards family, stickiness, and the calculus of $\chi_\text{st}$ are in the supplemental material~\cite{supp}.

%\section{Quantum $C_n$-Symmetrical Chaotic Billiards }
%\label{sec:quantum}

{\it Quantum $C_n$-Symmetrical Chaotic Billiards} -- We compute the energy eigenvalues using a boundary method~\cite{ver:1995} for a given billiard and evaluate the nearest neighbor spacing distribution $p(s)$~\cite{sto:2000}. The spectral statistics must be analyzed within symmetry-resolved subspaces for each value of $n$, since the Hamiltonian of a system with $C_n$ symmetry admits $n$ subspaces of eigenfunctions $\psi^{(m)}$ where $m$ is the eigenvalue of angular momentum along $z$ axis, $\hat{L}_z$. Recently, effects of well-defined angular momentum in spectral statistics were studied in~\cite{cas:2026}. For odd $n$, $m = \left (-\frac{n-1}{2}, \dots, -2, -1, 0, 1, 2, \dots, \frac{n-1}{2} \right )$, yielding an unique singlet subspace~($m=0$) and $(n-1)/2$ pairs of doublets~($m=\pm1, \pm2, \dots, \pm\frac{n-1}{2}$). For even $n$, one obtains $m = \left (-\frac{n-2}{2}, \dots, -2, -1, 0, 1, 2, \dots, \frac{n-2}{2}, \frac{n}{2} \right )$, leading to two singlet subspaces~($m=0$ and $m=n/2$), and $(n-2)/2$ pairs of doublets~$\left (\pm1, \pm2, \dots, \pm\frac{n-2}{2} \right )$. More details about the subspaces of well-defined angular momentum and the implementation of the boundary method are in the supplemental material~\cite{supp}.

\begin{figure*}[!htpb]
 \centering
 \includegraphics[width=2.0\columnwidth]{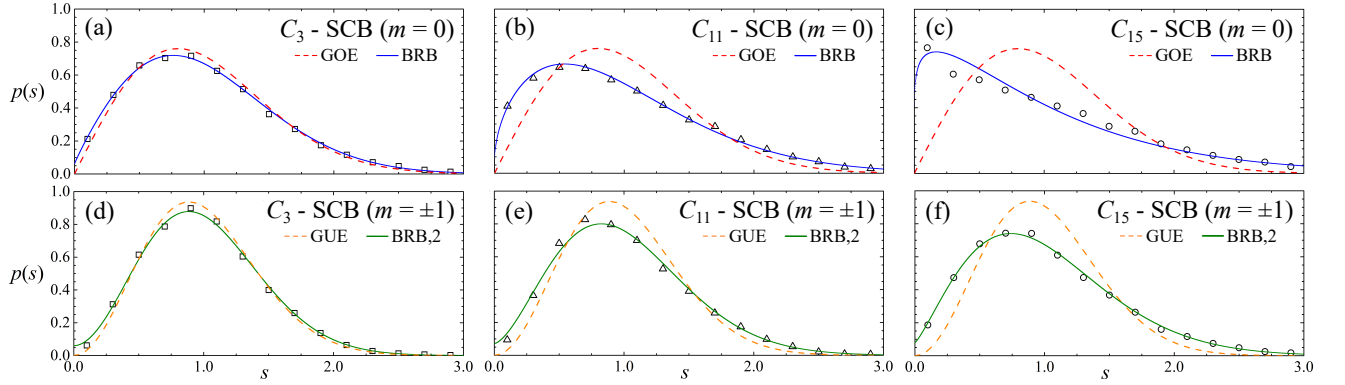}
 \caption{Nearest-neighbor level spacing distributions $p(s)$ for billiards with $r=0.25$ and symmetry orders $n=3, 11,$ and $15$. Panels (a)--(c): singlet subspace ($m=0$), showing deviations from GOE (red dashed) well fitted by the BRB distribution (blue solid). Panels (d)--(f): doublet subspace ($m=\pm1$), exhibiting deviations from GUE (orange dashed), accurately described by the BRB,2 distribution (green solid). Suma of fitted parameters: For $C_{3}$, $\beta_0= 0.92 \pm 0.02$ and $\gamma_1 = 0.97 \pm 0.02$. For $C_{11}$, $\beta_0= 0.54 \pm 0.02$ and $\gamma_1 = 0.78 \pm 0.03$. For $C_{15}$, $\beta_0= 0.15 \pm 0.03$ and $\gamma_1 = 0.61 \pm 0.01$}
 \label{fig:p_s_cn}
\end{figure*}

For $r=0.25$ and $n=3$, the billiard reduces to the LSS geometry. In our first analysis, we fix this radius and vary the symmetry parameter $n$. It is known that for $n=3$, the singlet spacing~($m=0$) distribution follows GOE statistics, while the spacing distribution for each member of the unique doublet pair~($m \pm 1$) follows GUE statistics. For each $p(s)$, we use 10,000 eigenvalues after discarding the first 1,000. Fig.~\ref{fig:p_s_cn} shows the behavior of the level spacing distributions for $n=3, 11$, and 15. Panels~(a)--(c) correspond to the singlet subspace~($m=0$), where we observe significant deviations of $p(s)$ from the GOE prediction~(red dashed line) for $C_{11}$ and $C_{15}$. These distributions are well described by the Berry--Robnik--Brody~(BRB) distribution with two parameters~\cite{pro:1994}: $\beta$, which quantifies the level repulsion and degree of localization in the singlet subspace, and $\rho_\text{st}$, a classical parameter corresponding to the fraction of sticky regions in phase space~(see Fig.~\ref{fig:billiards}).
\begin{multline}
 p_\text{BRB}(s) \e^{\rho_\text{st} s} = \\
 \frac{\rho_\text{st}^2}{\Gamma \left (\frac{1}{\beta+1} \right )} Q \left [ \frac{1}{\beta+1};a_\beta (\rho_\text{c} s)^{\beta+1} \right ] + \\
 [ 2 \rho_\text{st} \rho_\text{c} + (\beta+1) a_\beta \rho_\text{c}^{\beta+2}s^\beta ] \exp[-a_\beta(\rho_\text{c} s)^{\beta+1}],
 \label{eq:brb}
\end{multline}
where $a_\beta = \left [ \Gamma \left ( \frac{\beta+2}{\beta+1} \right ) \right ]^{\beta+1}$, and $Q(\kappa;x)$ is the Incomplete Gamma function. $\rho_\text{c}$ denotes the chaotic fraction of phase space and obeys the constraint $\rho_\text{st} + \rho_\text{c} = 1$. The solid blue lines in Fig.~\ref{fig:p_s_cn}(a)-(c) represent fits obtained by considering $\beta_m$ as a free parameter and $\rho_\text{st}$ extracted from values in Fig.~\ref{fig:billiards}.

For the doublet states, we select the subspace $m=1$, and Fig.~\ref{fig:p_s_cn}(d)-(f) display the corresponding $p(s)$ for billiards with $n=3,11$ and 15. Similarly to the singlet case, the doublet subspaces exhibit deviations from the GUE prediction~(orange dashed line). These deviations are well described by a distribution analogous to BRB, capturing the transition between Poisson and GUE statistics~\cite{ara:2021,ara:2024}. We refer to this distribution as BRB,2.
\begin{multline}
p_{\text{BRB,2}}(s) \e^{\rho_\text{st} s} = \\
\rho_\text{st} \rho_\text{c} b_{\gamma}^{\frac{1}{\gamma + 1}}\left(2 - \rho_\text{st} s \right) Q\left[ \frac{1+2\gamma}{1+\gamma};b_{\gamma}\left(\rho_\text{c} s \right)^{\gamma +1} \right] + \\
\bigg [ \rho_\text{st}^2\left(1+b_{\gamma}\rho_\text{c}^{\gamma+1}s^{\gamma+1}\right) + \\
(1+\gamma)\left(\rho_\text{c}^{\gamma+1}b_{\gamma}s^{\gamma} \right)^2 \bigg ] e^{-b_{\gamma}\left(\rho_\text{c} s \right)^{\gamma + 1}},
\label{eq:brb2}
\end{multline}
where $b_{\gamma} = \left[\Gamma\left(\frac{2\gamma+1}{\gamma+1}\right)\right]^{-(\gamma+1)}$. This distribution is also characterized by two parameters: $\gamma_{m}$, which quantifies the degree of localization and level repulsion for the symmetry classes of the doublets, and $\rho_\text{st}$. The solid green lines in Figs.~\ref{fig:p_s_cn}(d)--(f) correspond to fits using BRB,2 with just $\gamma_m$ considered as a free parameter.

\begin{figure}[!htpb]
 \centering
 \includegraphics[width=1.0\columnwidth]{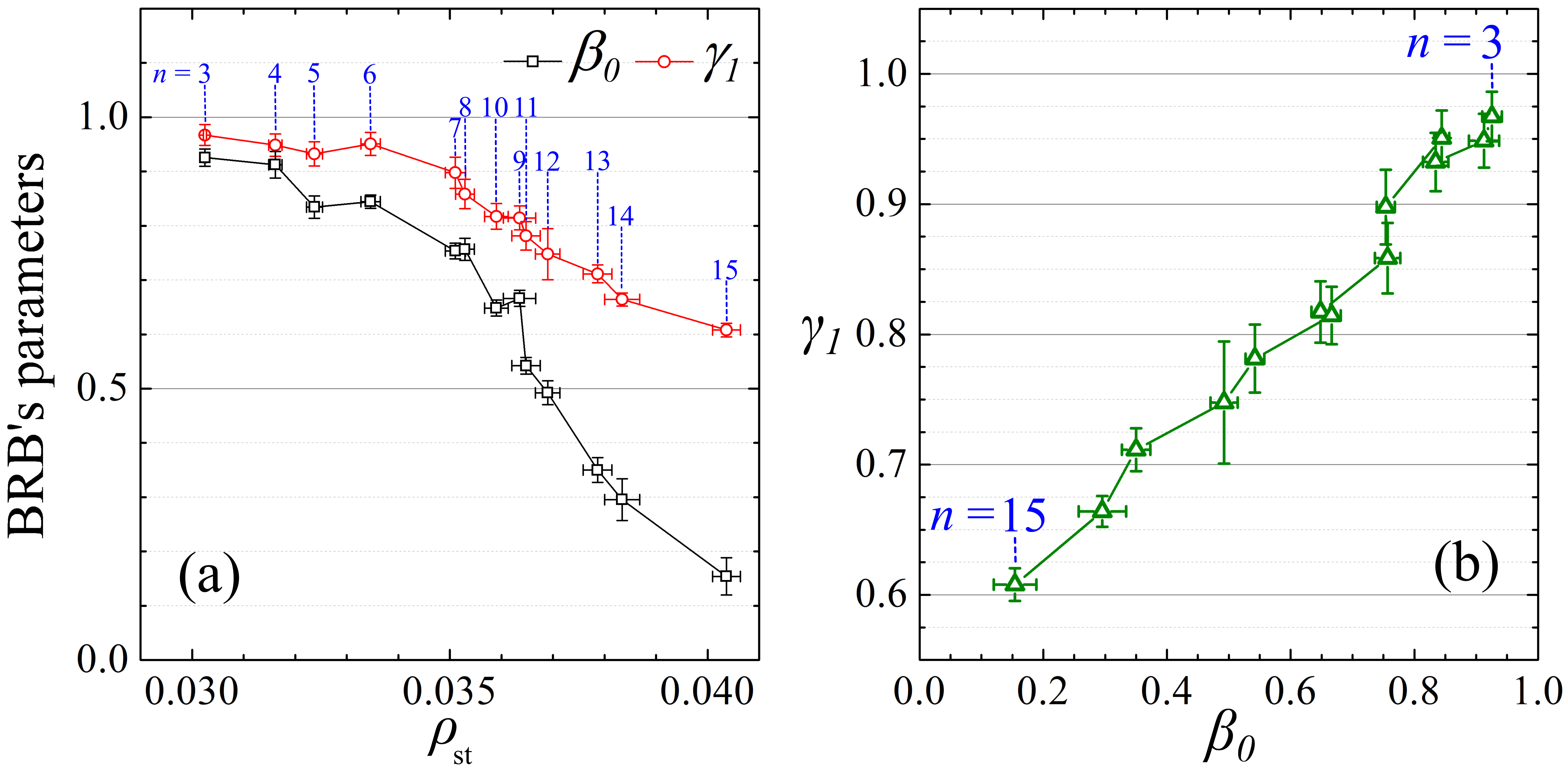}
 \caption{(a) Dependence of the fitting parameters $\beta_0$ (singlets, $m=0$) and $\gamma_1$ (doublets, $m=\pm1$) on the sticky phase-space fraction $\rho_{\mathrm{st}}$ for billiards with $n$ ranging from $3$ to $15$. Both parameters decrease monotonically with increasing $\rho_{\mathrm{st}}$, indicating growing deviations from GOE and GUE statistics, respectively. (b) $\gamma_1$ as a function of $\beta_0$, revealing a strong linear correlation (Pearson coefficient correlation $\simeq 0.99$).}
 \label{fig:beta_gamma}
\end{figure}

We highlight two main results at this stage: (i) increasing the symmetry parameter $n$ and the extent of sticky regions $\rho_\text{st}$ leads to progressively larger deviations from GOE and GUE distributions for singlets~($m=0$) and doublets~($m = \pm 1$), respectively. Even in a shear ergodic system; (ii) the simultaneous deviations in the level spacing distributions $p(s)$ for singlets and doublets in Fig.~\ref{fig:p_s_cn}(a) suggest correlations between different angular momentum $m$ within the same billiard. Fig.~\ref{fig:beta_gamma} reinforces the correlation hypothesis by showing a decreasing trend in both $\beta_0$ and $\gamma_1$ for a larger set of billiards as $\rho_{\mathrm{st}}$ increases, with the corresponding symmetry parameter $n$ ranging from $3$ to $15$. Fig.~\ref{fig:p_s_cn}(b) displays $\gamma_1$ as a function of $\beta_0$. The linear relationship between these parameters is quantified by the Pearson correlation coefficient, yielding approximately $0.99$. This value, being very close to $1$, indicates a strong positive linear correlation.

%\subsection{Level statistics in subspectra}
%\label{sec:level_sub}

\begin{figure*}[!htpb]
 \centering
 \includegraphics[width=2.0\columnwidth]{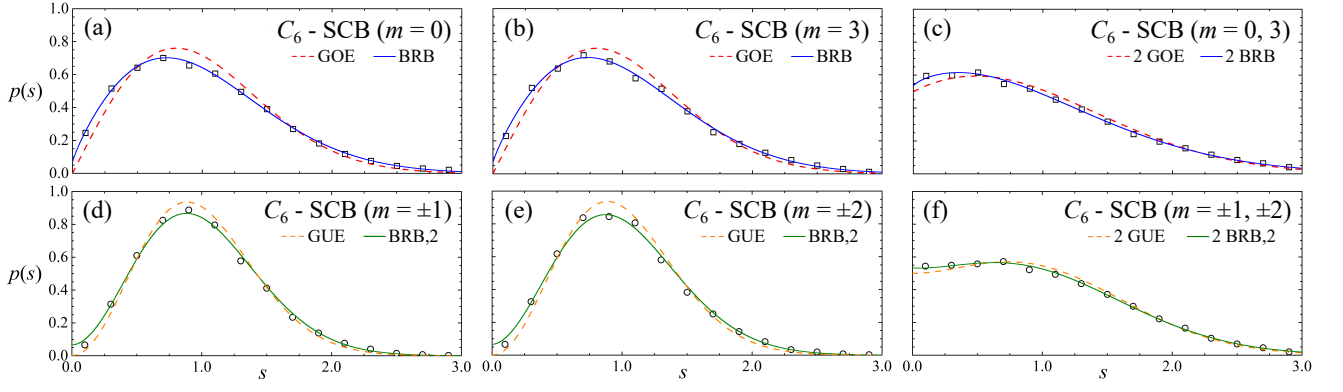}
 \caption{Nearest-neighbor spacing distributions $p(s)$ for a $C_6$-symmetric billiard ($\rho_{\mathrm{st}}=0.0334$). Panels (a),(b): singlet subspaces $m=0$ and $m=3$, exhibiting small deviations from GOE~(dashed lines), well captured by BRB fits~(solid lines) with compatible parameters $\beta_0=0.84\pm0.01$ and $\beta_3=0.86\pm0.02$. Panels (d),(e): doublet subspaces $m=\pm1$ and $m=\pm2$, showing analogous deviations from GUE~(dashed lines), accurately described by BRB,2~(solid lines) fits with $\gamma_1=0.95\pm0.02$ and $\gamma_2=0.92\pm0.02$. Panels (c),(f): superposition of singlet and doublet sequences, respectively, dashed lines denote GOE/GUE compositions, while solid lines correspond to BRB/BRB,2 compositions constructed from the fitted parameters, capturing the residual deviations and approaching Poisson statistics.}
 \label{fig:c6}
\end{figure*}

{\it Level statistics in subspectra} -- For even $n$, one can compare $p(s)$ of the singlet subspaces $m=0$ and $m=n/2$. For systems with fully chaotic dynamics and no stickiness, Refs.~\cite{li:2020,zha:2023b} showed that the quantization of billiards with $C_4$ symmetry yields two singlet subspaces whose level statistics follow the GOE prediction. In this context, panels~(a) and~(b) of Fig.~\ref{fig:c6} display analogous results for the $C_6$ symmetric billiard, with only small deviations from GOE in the $p(s)$ distributions of the singlet subspaces $m=0$ and $m=3$. Fits performed with the BRB distribution produce compatible parameters, $\beta_0=0.84 \pm 0.01$ and $\beta_3=0.86 \pm 0.02$. A similar behavior is observed between the doublet subspaces $m=\pm1$ and $m=\pm2$ in panels~(d) and~(e) of Fig.~\ref{fig:c6}, where simultaneous deviations from the GUE prediction are found. The fit parameters obtained from the BRB,2 distributions also exhibit compatible values, $\gamma_1=0.95 \pm 0.02$ and $\gamma_2= 0.92 \pm 0.02$. In all analyzed cases, $\rho_{\mathrm{st}}=0.0334$ was kept fixed.

\begin{figure*}[!htpb]
 \centering
 \includegraphics[width=2.0\columnwidth]{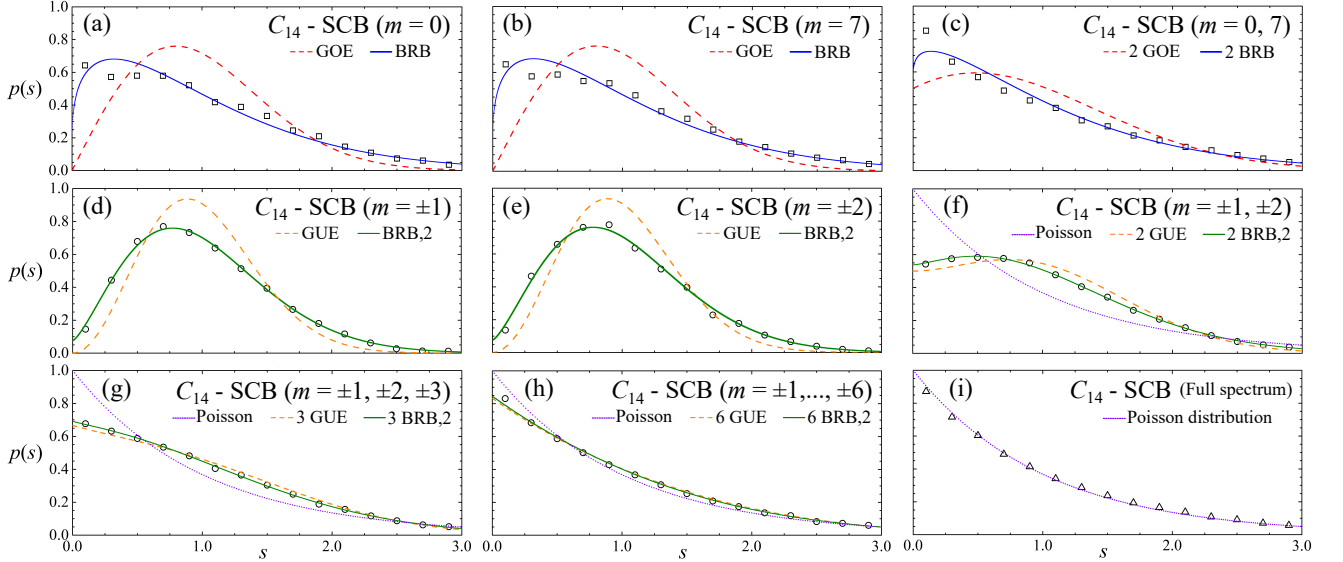}
 \caption{Nearest-neighbor spacing distributions $p(s)$ for the $C_{14}$ billiard ($\rho_{\mathrm{st}}=0.0383$). Panels (a),(b): singlet subspaces $m=0$ and $m=7$, showing simultaneous deviations from GOE~(dashed lines), well described by BRB~(solid lines) fits with compatible parameters $\beta_0=0.30\pm0.04$ and $\beta_7=0.29\pm0.04$. Panel (c): superposition of the two singlet sequences, the GOE composition~(dashed) fails to reproduce the data, while the BRB composition~(solid), constructed from the fitted parameters, captures the observed deviations, supporting correlations between singlet subspaces. Panels (d),(e): representative doublet pairs $m=\pm1$ and $m=\pm2$, showing similar deviations from GUE, well described by BRB,2 with compatible $\gamma$ parameters ($\gamma_1 = 0.66 \pm 0.01 $ and $\gamma_2 =0.67 \pm 0.02$;). Panels (f)--(h): superposition of an increasing number of doublet pairs with $\gamma_1$, $\gamma_2$, $\gamma_3 = 0.63 \pm 0.02$, $\gamma_4 = 0.65 \pm 0.01$, $\gamma_5 = 0.64 \pm 0.02$ and $\gamma_6 = 0.66 \pm 0.01$, illustrating the rapid approach to Poisson statistics; compositions of GUE (orange dashed) and BRB,2 (green solid) become nearly indistinguishable for six sequences. Panel (i): full spectrum, including singlets ($m=0,7$) and all doublets ($m=1,\ldots,6$), yielding a Poisson distribution (purple dashed), despite non-GOE/GUE statistics in the individual subspectra.}
 \label{fig:c14}
\end{figure*}

The superposition of many eigenenergy sequences leads to the Poisson distribution~\cite{ber:1984,guh:1998,abu:2009}, and this behavior can be explicitly observed in the $C_n$-SCB billiards. Fig.~\ref{fig:c6}(c) displays the spacing distribution $p(s)$ for the superposition of the two singlet sequences, together with the curves corresponding to the composition of two GOE distributions~(red dashed line) and two BRB distributions~(solid blue line). The BRB superposition accounts for the small deviations observed in the individual spectra. The corresponding curve was constructed using the fit parameters obtained in panels~(a) and~(b). Similarly, panel~(f) of Fig.~\ref{fig:c6} shows the superposition of the doublet subspaces $m=\pm1$ and $m=\pm2$, together with the composition of two GUE distributions~(orange dashed line) and two BRB,2 distributions~(solid green line). Although subtle, the BRB,2 superposition captures the deviations present in the isolated spectra of the different doublet subspaces. This curve was plotted using the parameters $\gamma_1$ and $\gamma_2$ previously obtained in panels~(d) and~(e). Details on the superpositions of GOE, GUE, BRB, and BRB,2 statistics are provided in the Supplemental Material~\cite{supp}. For the $C_6$ billiard, one observes a correlation between the two singlet subspectra, as both display subtle deviations from the GOE prediction. These deviations become more pronounced in the $C_{14}$ billiard, where Figs.~\ref{fig:c14}(a) and (b) show that the level-spacing distributions for $m=0$ and $m=7$ depart simultaneously from GOE, reinforcing the hypothesis of correlations between different singlet subspaces. The resulting distributions are well described by the BRB distribution, with compatible fitting parameters $\beta_0 = 0.30 \pm 0.04$ and $\beta_7 = 0.29 \pm 0.04$. The stickiness fraction for the $C_{14}$ billiard were computed as $\rho_{\mathrm{st}} = 0.0383$. In panel Fig~\ref{fig:c14}(c), the superposition of the subspaces $m=0$ and $m=7$ exhibits a significant deviation when compared with the composition of two GOE distributions. The superposition of two BRB distributions satisfactorily captures this behavior.

For the same billiard, there are six distinct doublet pairs, which likewise exhibit larger deviations from GUE when compared with the two doublet pairs analyzed in the $C_6$ billiard. All six pairs display similar level statistics, and the BRB,2 distribution provides an accurate description of these results. As representative examples of the correlation among the curves, Figs.~\ref{fig:c14}(d) and (e) show the distributions for the pairs $m=\pm1$ and $m=\pm2$. The corresponding values of $\gamma$ are mutually compatible across the six pairs and are listed in the caption of the same figure. In order to understand how the spacing distribution of the full spectrum emerges, Fig.~\ref{fig:c14} displays the contributions arising from the combinations of different doublet pairs. Panels \ref{fig:c14}(f)--(h) show a rapid approach to the Poisson distribution as additional doublet pairs are included in the superposition statistics. It is worth emphasizing that the composition of GUE curves (orange dashed line) and the composition of BRB,2 curves become nearly indistinguishable in the case of six sequences, as observed in panel~\ref{fig:c14}(h). Finally, Fig.~\ref{fig:c14}(i) displays the spacing distribution of the full spectrum, including the singlets $m=0$ and $m=7$, as well as the doublet sectors $m=1,\ldots,6$. The resulting distribution (black triangles) is well described by the Poisson distribution (purple dashed line). Although the Poisson distribution is obtained when the full energy spectrum of the billiard is analyzed, one might be led to conclude that it arises from a superposition of GOE or GUE sequences, which is in fact not the case. The most appropriate technical procedure is to examine the behavior of the individual subspectra and then relate their properties to the underlying classical dynamics of the system.

%\section{Conclusions}
%\label{sec:conc}

{\it Conclusions} -- We have numerically investigated the quantum spectral properties of a family of fully chaotic billiards with discrete rotational symmetry $C_n$, whose geometry generalizes the Leyvraz, Schmit and Seligman construction \cite{ley:1996}. Although the classical dynamics remains ergodic, the presence of stickiness in phase space produces clear nonuniversal effects in the symmetry-resolved spectra. In particular, we showed that the spacing distributions in singlet and doublet subspectra systematically deviate from the GOE and GUE, respectively, and that these deviations become stronger as the sticky phase space fraction $\rho_{\mathrm{st}}$ increases. The results are accurately described by the BRB and BRB,2 formulas~\cite{ara:2021,pro:1994}, establishing a quantitative connection between spectral fluctuations and classical trapping effects.

A central result concerns the emergence of correlations between distinct subspaces of angular momentum $m$. For even $n$, the two singlet subspaces~($m=0$ and $n/2$) exhibit compatible values of $\beta$, indicating simultaneous deviations from GOE as $\rho_\text{st}$ increases. Among the doublet pairs, the simultaneous deviations from GUE yield compatible values of $\gamma$ for different subspaces~($m=\pm1, \pm2,\ldots$). Moreover, we observe that correlations also arise between $m=0$ and $m=\pm1$, as increasing $\rho_\text{st}$ leads to simultaneous variations in both $\beta_0$ and $\gamma_1$~(Fig.~\ref{fig:beta_gamma}). In general, $\beta$ decreases more rapidly than $\gamma$ as $\rho_\text{st}$ increases, indicating that states with angular momentum $m=0$ and $n/2$~(invariant under time reversal) are less localized than those with $m=\pm1, \pm2,\ldots$~(not invariant under time reversal). Finally, by analyzing the superposition of multiple subspectra, we showed that intermediate BRB-type statistics successfully capture the residual deviations that survive when only a few sequences are combined. As additional sequences are superposed, the spacing distribution rapidly approaches the Poisson limit \cite{ber:1984,guh:1998,abu:2009}. In this process, the degenerated sequences play a dominant role, since the number of singlet subspaces alternates only between one and two, whereas the number of doublet pairs grows with $n$. This demonstrates that the statistical behavior of the complete spectrum may hide the nonuniversal information encoded in the symmetry-resolved subspaces.

Our results establish chaotic billiards with $n$-fold symmetry as a valuable framework for investigating the interplay between discrete symmetries, time-reversal invariance, and classical stickiness in quantum chaos. We focused on relatively low values of the symmetry parameter $n$, for which the deviations from the GOE and GUE curves can already be clearly identified. A natural extension of the present results is the investigation of the $C_n$-SCB family at higher energies and for larger symmetry orders. Besides, others' future directions include the study of long-range spectral correlations~\cite{boh:1984}, semiclassical periodic-orbit descriptions~\cite{gut:1971}, localization measures ~\cite{bat:2013,bat:2019,loz:2021,ore:2026}, and experimental realizations in microwave, acoustic, or graphene-based billiards with engineered rotational symmetry~\cite{zah:2023,zhan:2023}.

%\begin{acknowledgments}

{\it Acknowledgments} -- Valuable discussions with F. M. de Aguiar and Washington F. dos Santos are gratefully acknowledged. We are also grateful for the computational resources of LMCR~(Brazilian Agency CNPq under grant no. 404592/2025-2) from the Department of Physics at Universidade Federal Rural de Pernambuco.

%\end{acknowledgments}

%\nocite{*}

\bibliography{refs_2026_RBdoCarmoTAraujoL_Quantum_nFoldSymmetry.bib}% Produces the bibliography via BibTeX.

\end{document}